# Topological Woodward-Hoffmann classification for cycloadditions in polycyclic aromatic azomethine ylides

Juan Li[1,#], Amir Mirzanejad[5,#], Wen-Han Dong[2,#], Kun Liu[3], Marcus Richter[3], Xiao-Ye Wang[6,7], Reinhard Berger[3], Shixuan Du[2], Willi Auwärter[4], Johannes V. Barth[4], Ji Ma[3], Klaus Müllen[6], Xinliang Feng[3], Jia-Tao Sun[8*], Lukas Muechler[5], Carlos-Andres Palma[2,9*]

[1]*Advanced Research Institute for Multidisciplinary Science, Beijing Institute of Technology, 100081 Beijing, China.*

[2]*Institute of Physics & University of Chinese Academy of Sciences, Chinese Academy of Sciences, 100190 Beijing, China.*

[3]*Chair for Molecular Functional Materials, Center for Advancing Electronics Dresden (cfaed), Faculty of Chemistry and Food Chemistry, Dresden University of Technology, Mommsenstr. 4, 01062 Dresden, Germany.*

[4]*Physics Department E20, Technical University of Munich, James-Franck-Str. 1, 85748 Garching, Germany.*

[5]*Department of Chemistry, Penn State University, 108 Chemistry Building, 16802 University Park, United States.*

[6]*Max Planck Institute for Polymer Research, Ackermannweg 10, 55128 Mainz, Germany.*

[7]*State Key Laboratory of Elemento-Organic Chemistry, College of Chemistry, Nankai University, 300071 Tianjin, China.*

[8]*School of Integrated Circuits and Electronics, MIIT Key Laboratory for Low-Dimensional Quantum Structure and Devices, Beijing Institute of Technology, 100081 Beijing, China.*

[9]*Department of Physics & IRIS Adlershof - Humboldt-Universität zu Berlin, 12489, Berlin, Germany*

**Abstract**

The study of cycloaddition mechanisms is central to the fabrication of extended $sp^2$ carbon nanostructures. Reaction modeling in this context has focused mostly on putative, energetically preferred, exothermic products with limited consideration for symmetry allowed or forbidden mechanistic effects. Here, we introduce a scheme for classifying symmetry-forbidden reaction coordinates in Woodward-Hoffmann correlation diagrams. Topological classifiers grant access to the study of reaction pathways and correlation diagrams in the same footing, for the purpose of elucidating mechanisms and products of polycyclic aromatic azomethine ylide (PAMY) cycloadditions with pentacene–yielding polycyclic aromatic hydrocarbons with an isoindole core in the solid-state and on surfaces as characterized by mass spectrometry and scanning tunneling microscopy, respectively. By means of a tight-binding reaction model and density functional theory (DFT) we find topologically-allowed pathways if a product is endothermic, and topologically-forbidden if a product is exothermic. Our work unveils topological classification as a crucial element for reaction modeling for nanographene engineering, and highlights its fundamental role in the design of cycloadditions in on-surface and solid-state chemical reactions, while underscoring that exothermic pathways can be topologically-forbidden.

**Introduction**

Cycloaddition reactions[1,2] are cornerstones in carbon nanomaterial engineering. Early examples include Diels-Alder[3,4] [4+2] and Prato-type[5] [3+2] cycloadditions in solution environments, and Huisgen-type[6] [3+2] as well as related [2+2] Bergman[7,8] cyclization on surfaces. With the advent of nanographene synthesis[9,10-12], a new chapter in organic chemistry has opened up, seeking a modular, highly selective, and high-yield cycloadditions[13-25] of extended conjugated macromolecules at interfaces without byproducts.[26] This endeavor embodying click-chemistry[27] has notably driven the adaptation of a large variety of organic reactions at interfaces, including Ullmann coupling, Glaser coupling and polycondensations. Recently, we have shown that polycyclic aromatic azomethine ylides (PAMYs, **Fig. 1a**) can be employed to form diaza-hexabenzocoronenes and N-containing polycyclic aromatic chains[21,28] in the solid-state and on surfaces, opening an avenue to cycloaddition polymerizations for extended polycyclic aromatic hydrocarbons (PAHs) or related nanographenes. In solution, the PAMY precursor, namely 8*H*-isoquinolino[4,3,2-*de*]phenanthridin-9-ium tetrafluoroborate (DBAP salt, **1**, see Methods and SI), undergoes selective 1,3-dipolar [3+2] cycloaddition to electron-deficient dipolarophiles yielding N-containing PAHs.[29,30] Towards the engineering of extended N-containing PAHs on-surface and in the solid-state, a mechanistic, broadly accessible understanding of the chemical reactions and pathways accessible to PAMY and similar cycloadditions is desirable.[31]

In on-surface and solid-state thermochemistry, chemical reaction modeling beyond adiabatic energetic diagrams (e.g. nudge elastic band method for determining transition structure and energy barrier[32]) is rarely explored since the potential energy surfaces of ground states are often assumed to follow the noncrossing rule[33] or void of electronic state hopping. Because of this limitation, it remains often unclear in the literature whether cyclization and dehydrogenation reactions are symmetry-allowed, that is, occur adiabatically or otherwise. Symmetry-forbidden reactions, that is, reactions in which molecular orbitals cross, are indicative of nonadiabatic dynamics that are known to play a fundamental role in PAH synthesis and cycloadditions[34,35]. Several models are available to attempt to formally define forbidden nonadiabatic reaction pathways from on-surface reaction mechanisms. Woodward and Hoffmann (WH)[36-40] attribute the reactivity of a chemical reaction to the atomic orbital symmetries under adiabatic conditions. The WH rules provide qualitative selection rules for pericyclic cycloaddition reactions[41,42-45] whereby molecular orbital symmetry and crossings are commonly treated by means of Fukui reaction theory[46], Marcus theory of electron transfer and the surface hopping method[47]. More accurate mechanistic predictions relying on the Born-Oppenheimer[48] approximation, are challenging when dealing with more than one reaction pathway in the presence of nonadiabatic quantum effects such as tunneling and surface hopping. These effects make it difficult to validate qualitative WH rules. Yet the WH approach remains a powerful well-known concept for reaction engineering. In this regard, quantitative and chemically-intuitive WH visualization tools treating ionic, radical and pericyclic reactions on the same footing would be desirable for the rapid prototyping of PAH reactions, and the study of nonadiabaticity. During the last decades, the classification of orbital crossings[49-51] aided by differential geometry and algebraic topology has emerged as a promising method to study nonadiabaticity in condensed matter physics.[52-59] Topological physical properties[60] in quantum matter and metamaterials can now be engineered with an extraordinary level of sophistication, aided by the interplay between effective ('toy') models, *ab initio* calculations and experiments. Such methodology has been rarely employed for the study of chemical phase transitions and corresponding reactions. Recently, the concept of topology classification for mirror-symmetric reaction pathway models to study reactions by means of topological invariants was introduced[61,62], whereby the reactions with distinct topologically classifiers are adiabatically forbidden (such as the [2+2] thermal reaction of two ethylene molecules)[61,62]. Topological classifications could epitomize a turning point in cycloaddition reaction engineering, especially on-surface, where reactions are surface templated and highly symmetric. Particularly, such topological models expand and unify the WH–Fukui approach: They enforce the geometrical symmetry and concertedness of reaction pathways to summarize and formalize chemical notions, simplifying reaction interpreting and rational design. Additionally, they illustrate that reaction coordinates can be mathematically defined to study nonadiabaticity and (non-interacting) orbital intersections from a topological standpoint.[63,64]

Here, we study the solid-state and on-surface cycloaddition of a PAMY precursor and pentacene to yield internally N-containing PAH with a tetracenoisoindole core (**Fig. 1a**) as characterized by ultra-high vacuum (UHV) scanning tunneling microscopy (STM) and matrix-assisted laser desorption-ionization mass spectrometry (MALDI-MS).[65] By investigating the frontier orbital symmetries of gas-phase reaction pathways by means of intrinsic reaction coordinates which are assumed conclusive for the study of the on-surface and in the solid-state reactions, we describe the [3+2] reaction (**Fig. 1b**) between singlet diradicaloid PAMY (rPAMY) and pentacene and show that its de-aromatization pathway is adiabatic and therefore thermally allowed in the gas phase. We formally classify the WH rules via topological invariants, extending the recently proposed topological classification[61] to a tight-binding *Hückel reaction* model combining first-principles calculations (**Fig. 1c**). Different from the frontier orbital model, our topological WH model differentiates the allowed, concerted adiabatic pathways from the nonadiabatic, crossings by a $\mathbb{Z}$-classified topological invariant $C(t) = N_+(t) - N_-(t)$, where $N_+(t)$ ( $N_-(t)$ ) is the number of mirror symmetric (antisymmetric) molecular orbitals (MOs) in all occupied MOs (**Fig. 1c**). We find that singlet diradicaloid PAMY lateral addition to acenes is endothermic but topologically WH allowed, while central ring addition is exothermic but topological forbidden. Our work introduces a methodological and theoretical approach for the study of cycloaddition selectivity, particularly PAMY reactions which are relevant in development of N-containing PAHs as substrates for N-doped nanographenes, spin-chains[66,67], metal-free catalysis[68,69] and sensors[70,71].

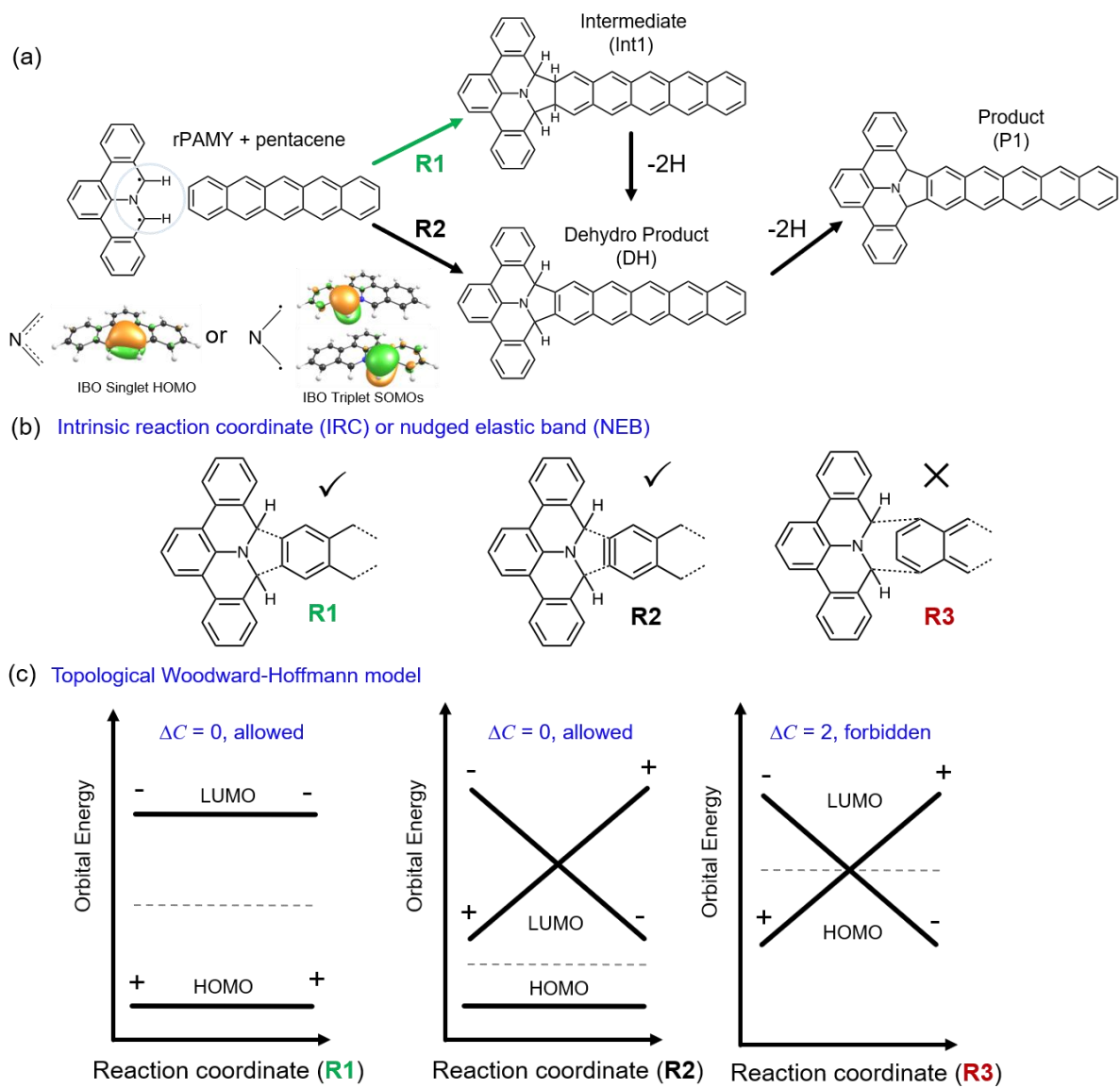

**Fig. 1 | Singlet diradicaloid PAMY cycloadditions and their symmetry-allowed or forbidden pathway classification as a topological obstruction** (**a**) The reaction scheme of diradicaloid PAMY (rPAMY) with pentacene posits some possible reaction pathways, one associated with pentacene de-aromatization intermediate (Int1) and the other without a dehydrogenated intermediate implying pentacene aromaticity-conserving pathways (R2). The rPAMY triplet lies 25 kcal mol$^{-1}$ above the singlet and hence only the singlet reactive species is considered in this work. The intrinsic bond orbital (IBO) DFT analysis identifies the ground state as a singlet resonant structure of the diradical triplet state. (**b**) Three cycloaddition pathways R1, R2, and R3 between singlet rPAMY and pentacene by means of intrinsic reaction coordinates or nudged elastic band. (**c**) Topological Woodward-Hoffmann classification. Woodward–Hoffmann (WH) frontier orbital rationalization of allowed or forbidden reactions the where "+" ("–") denotes an even (or odd) MO with respect to a symmetry reference (see SI **Fig. S1, S2**). The dashed line separates occupied (HOMO) from unoccupied orbitals (LUMO). In topological classifications, explicit parameters for the analytical understanding of reaction coordinates, isomeric reaction pathways in a formal mathematical framework for the topological classification of occupied frontier orbital intersections are explored, beyond WH rules alone.

**Cycloaddition between rPAMY and pentacene: MALDI-MS and STM**

Unlike homocoupling solution reactions[28], previous heterocoupling solid-state studies[72] employing DBAP salt have not identified key intermediates which unambiguously evidence cycloaddition reaction pathways. Therefore, we set to investigate on-surface and solid-state intermediates and products between the DBAP and pentacene in **Fig. 2** employing mass spectrometry (MS) and scanning tunneling microscopy (STM). Upon heating to 250 °C a 1:1 solid-state mixture of pentacene and DBAP salt, a peak of $m/z = 543.198$ is identified in the matrix-assisted laser desorption/ionization (MALDI) mass spectrum (**Fig. 2a**), assigned to the heterocoupling product with a hydrogenated tetracenoisoindole core, together with the expected competing homocoupling products[28]. A tentative structure for the product is **DH** (**Fig. 1a**, **2a**) ensuing from a hydrogenated tetracenoisoindole core and partial dehydrogenation of pentacene. Such product could occur following pathway **R1** (**Fig. 1a**) as a probable adiabatic cycloaddition mechanism. A direct, unknown pathway **R2** could occur but is less plausible. These pathways consider that a hydrogen of the unreactive species of DBAP is removed, to form diradicaloid PAMY (rPAMY, **Fig. 1a**), which has been previously characterized on surfaces[28,72]. The diradicaloid term is used to address both potential spin states of rPAMY, whereby the singlet and triplet frontier orbitals are depicted in **Figure 1a** with intrinsic bond orbital (IBO) analysis. It is worth noting that triplet diradical PAMY is 25 kcal mol$^{-1}$ higher in energy, which is around 8 kcal mol$^{-1}$ above than the highest energy barrier found in this work, and therefore not considered as the reactive species.

When mixing 2.2 equivalents of DDQ in the solid-state reaction to further dehydrogenate the DH intermediate (see Methods), a fully dehydrogenated product, tentative **P1**, bearing a tetracenoisoindole core is found (**Fig. 2b**). The energy diagram in **Fig. 2e** depicts that the expected heterocoupling intermediate product of the R1 pathways TS1 is 13.6 kcal/mol above the reactants, whereby the intermediate with two detached hydrogens (DH) is more stable than the final product **P1**, consistent with the MALDI result.

LT STM measurements on the Ag(100) surface (**Fig. 2c** and **2d**) were carried out to further investigate surface-confined reaction products. Upon sublimation of pentacene and DBAP onto Ag(100) and annealing treatment at around 400 °C, the major product of the reaction is identified as bearing mirror symmetry with the 2,3 position of pentacene as studied in **Fig. 3**. Metal substrates are known to catalyze dehydrogenation, such that the reaction on Ag(100) is comparable to the solid-state reaction with DDQ and the reaction product is inferred as the expected final product **P1** with a tetracenoisoindole core. Moreover, the three-dimensional structure of **Int1**, with an angle of 125° between the PAMY and pentacene, strongly biases the reaction towards planarization through partial dehydrogenation products **DH** and **P1**. An additional minor product **P2** derived from the pentacene molecules reacting on the 1,2 positions is observed (**Fig. S14**).

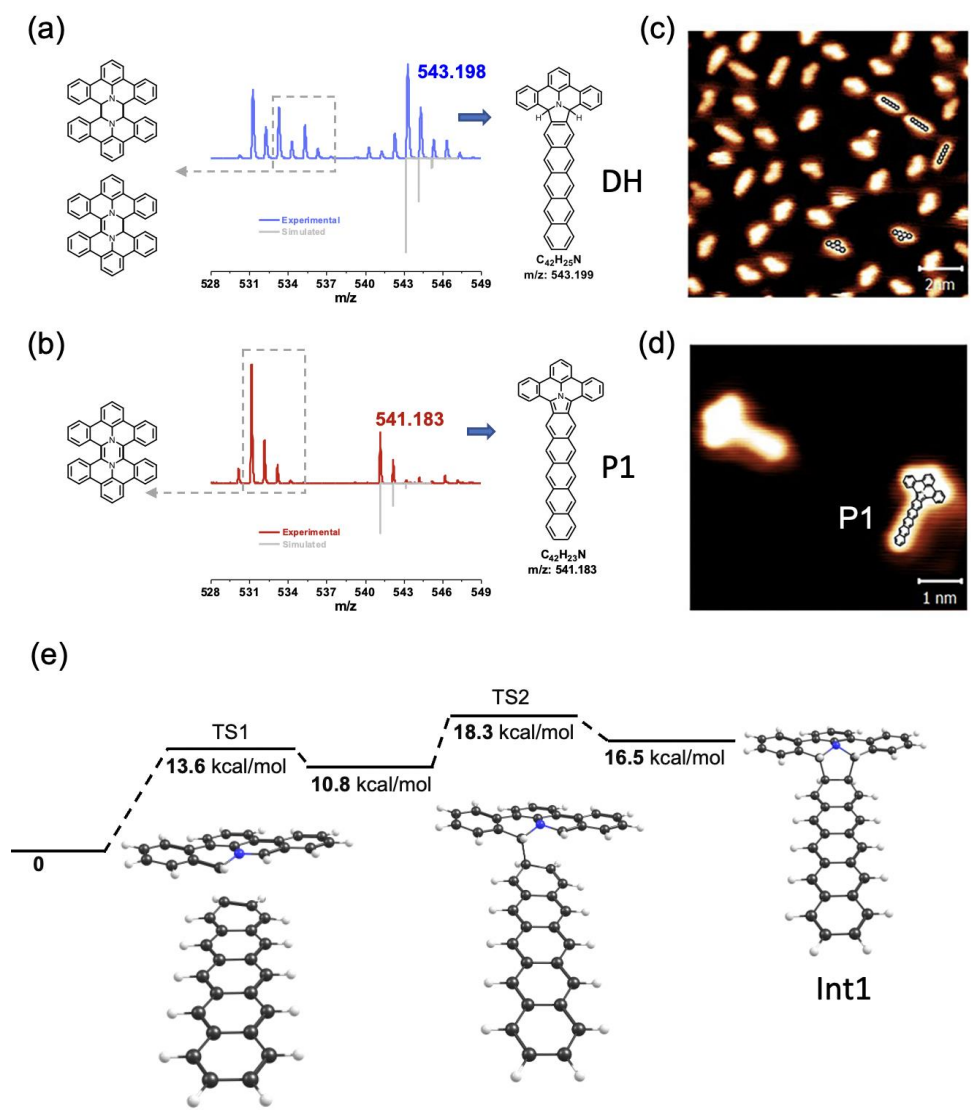

**Fig. 2 | Reaction of rPAMY precursor and pentacene involving de-aromatization of pentacene**. (**a**) MALDI-TOF mass spectra of the product from solid-state synthesis at 250 °C with DBAP salt and pentacene (1:1). (**b**) MALDI-TOF mass spectra from solid-state synthesis at 250 °C with DBAP salt, pentacene and DDQ (1:1:2.2). The competing products from rPAMY cycloaddition dimerization, diaza-based PAMY dimers, are also shown. (**c**) STM survey of the DBAP and pentacene reaction after simultaneous evaporation of precursor and pentacene on Ag(100). Scanning parameters: $I_t$ = 30 pA, $V_s$ = 300 mV. (**d**) High-resolution STM of the reaction product **P1** on Ag(100) after on-surface synthesis at ~ 400 °C. Scanning parameters: $I_t$ = 30 pA, $V_s$ = 300 mV. (**e**) IRC pathway and broken-symmetry DFT energy profile for the adiabatic **R1** pathway without substrate, showing the endothermic formation of the intermediate Int1 ($C_{42}H_{23}N$) product, which may further dehydrogenate into **DH** or **P1** (cf. **Fig. 1a**, **Fig. 2**).

**Cycloaddition between rPAMY and pentacene: Intrinsic reaction coordinate calculations**

To study isomeric reaction pathways, we focus on the cycloaddition between singlet diradicaloid PAMY (rPAMY), and pentacene by broken-symmetry DFT calculations. **Figure 2e** depicts the energy profile obtained from intrinsic reaction coordinate (IRC) calculation of the **R1** proposed reaction pathway leading to **P1** via on-surface dehydrogenation as detailed in supporting information. There are at least two additional plausible reaction pathways (**R2**–**R3**) which could explain the on-surface and solid-state data (**Fig. 3**). The calculated frontier orbitals of the starting configuration can be defined as symmetric (S) or antisymmetric (AS) with respect to the molecular mirror plane. These symmetries are conserved in the endothermic **R1** [3+2] cycloaddition IRC pathway (**Fig. 3a**, **b**) and therefore WH allowed. In pathway **R2**, which considers the cycloaddition reaction with dehydrogenated pentacene, the symmetry of the highest occupied molecular orbital is conserved and hence equally WH allowed. Here, attempts to locate a transition structure via IRC were unsuccessful, whereas nudge-elastic band identifies the reaction as barrierless and highly exothermic (**Fig. 3c**, **d**). Finally, the solid-state reaction could admit an isomer of product **P1** through pathway **R3**, in which orbitals symmetries are changed. Accordingly, the transition structure search fails to locate a concerted [3+4] cycloaddition mechanism, despite the **Int2** being exothermic (**Fig. 3e**, **f**). Apart from the concerted mechanism being WH forbidden, the pathway towards **Int2** is obstructed through hydrogen migration, see hydrogen in **Fig. 3f**. Further generalization to acenes in **Table S1** confirms that the reaction of rPAMY with acenes is a type-I HOMO-controlled [3+2] cycloaddition[44] since rPAMY is the electron donor, in line with suggested mechanisms for high-temperature on-surface and in-solution rPAMY dimerization[28] and a recently reported heterocycloaddition, the polymerization of cyano-rPAMY[72] to be further explored below. Classifying forbidden and allowed reaction pathways according to WH rules, offers qualitative criteria for the design of reaction pathways[73], yet a formal, topological classification could qualitative convey the symmetries and obstructions involved in cycloadditions.

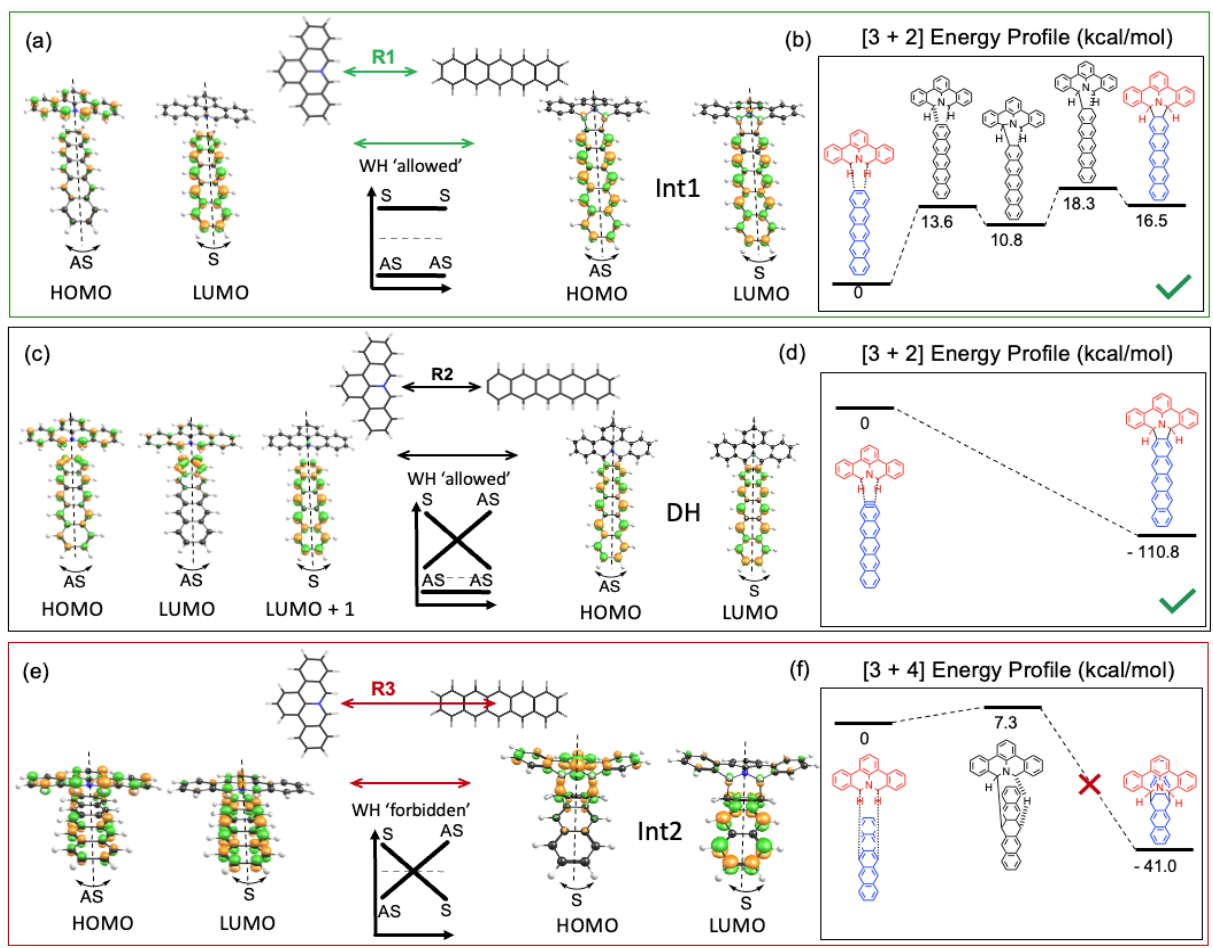

**Fig. 3 | Woodward-Hoffmann exploration of reaction pathways for rPAMY + pentacene cycloadditions which are plausible in the solid-state or on-surfaces. (a, b)** Endothermic PAMY+pentacene **R1** [3+2] intrinsic reaction coordinate (IRC) cycloaddition pathway to the intermediate product (Int1) towards fully dehydrogenated **P1**. The frontier orbitals are also depicted. The pathway **R1** takes place in a non-concerted manner. **(c, d)** For comparison the exothermic [3+2] cycloaddition pathway with an aryne moiety in pre-dehydrogenated pentacene is purely concerted and allowed. **(e, f)** The IRC exploration of exothermic PAMY+pentacene **R3** pathway, plausible in the solid-state, shows that a hydrogen migration reaction takes place in lieu of the WH forbidden, obstructed cycloaddition reaction towards **Int2**.

**Adiabatic vs. nonadiabatic pathways: Topological classification**

A symmetry forbidden WH pathway can be defined as orbitals crossing, and this crossing or obstruction topologically classified by a non-zero change of topological invariants. The topological invariants classify the discontinuities between eigenvector spaces of reaction matrices (see SI **Fig. S1** block diagonalization *Hückel reaction* in **Fig. 4** and topological classification). Mirror-symmetric (i.e. mirror symmetry-enforced) *Hückel reaction* models for the reaction between rPAMY and hydrocarbons can be constructed as depicted in **Fig. 4**. As an extended formal framework for WH rules, topological classification models can handle pericyclic and radical cycloadditions[74-76] and are highly useful for visualizing plausible reaction pathways with enforced reaction symmetry (mirror plane depicted in **Fig 4a**). Compared with classic WH diagrams, this model grants explicit access to the quantitative exploration of concerted cycloadditions which conserve the matrix symmetry by means of tunable bonding strengths. The model could be extended, in principle, to biradical cases[34,35] and complex geometrical transformations of molecules, provided certain symmetries are included.

In a *Hückel reaction*, we construct a general 6 × 6 or 8 × 8 *Hückel matrix* of the four-atom [3+2] or [3+4] representation of the reaction between the *p* orbitals of a rPAMY fragment and butadiene or benzene in **Fig. 4a, b** or **4c-e**, respectively. **Fig. 4e** shows 8 × 8 *Hückel matrix* for the forbidden [3+4] **R3**-like cycloaddition. It shows how to the key quantifying reaction coordinate t in the *Hückel reaction* model is inversely related to the distance between the reactants t = 1/d and crossings are identified through discontinuities in the frontier orbitals and their symmetries. Here, we assume that the valence electrons of the nitrogen are bonded to the carbons with strength *a*, and the onsite energy *p1 (p1\*)* differs from *p2* due to the heteroatom effect. To enforce symmetry and account for both *p* electrons of nitrogen, we consider a virtual *sp²* nitrogen where the lone pair occupies a degenerate pair of $p_z$ and $p_z^*$ orbitals, thereby virtually increasing the order of the cycloaddition from a [3+n] ring to a [4+n] ring. The two reactants interact by a single variable parameter t < 0 with |t| as the reaction coordinate. Forbidden pathways in the model indicated by crossings of molecular orbitals that are ℤ-classified by the topological invariant $C(t) = N_+(t) - N_-(t)$. (see **Fig. S2** and previous section). The topological analysis of the frontier orbital evolution in **Fig. 4b, d** (see details in **Fig. S2**) shows that the diradicaloid reaction is allowed with benzene ($\Delta C = 0$), but forbidden with the 1,4-positions of butadiene ($\Delta C = 2$), where $\Delta C = C_{react} - C_{prod}$ is the difference of the topological invariants between reactants and products. **Figure S4, S5** provides details of how to obtain the parameters by fitting the *Hückel*-like parameters with DFT orbitals. The tight-binding *Hückel reaction* **R3**-like pathway is extended with DFT calculations in **Figure 4f, g**. The DFT reaction pathway is only approximately forbidden, as strict eigenstate crossings seldom occur beyond tight-binding models. In such cases, Green function formalism can be employed to classify topological crossings in highly correlated reaction pathways[77,78]. In summary, cycloaddition WH reaction matrix models help formalize WH rules and vastly extend the scope of the engineering of adiabatically forbidden or allowed radicaloid reaction pathways.

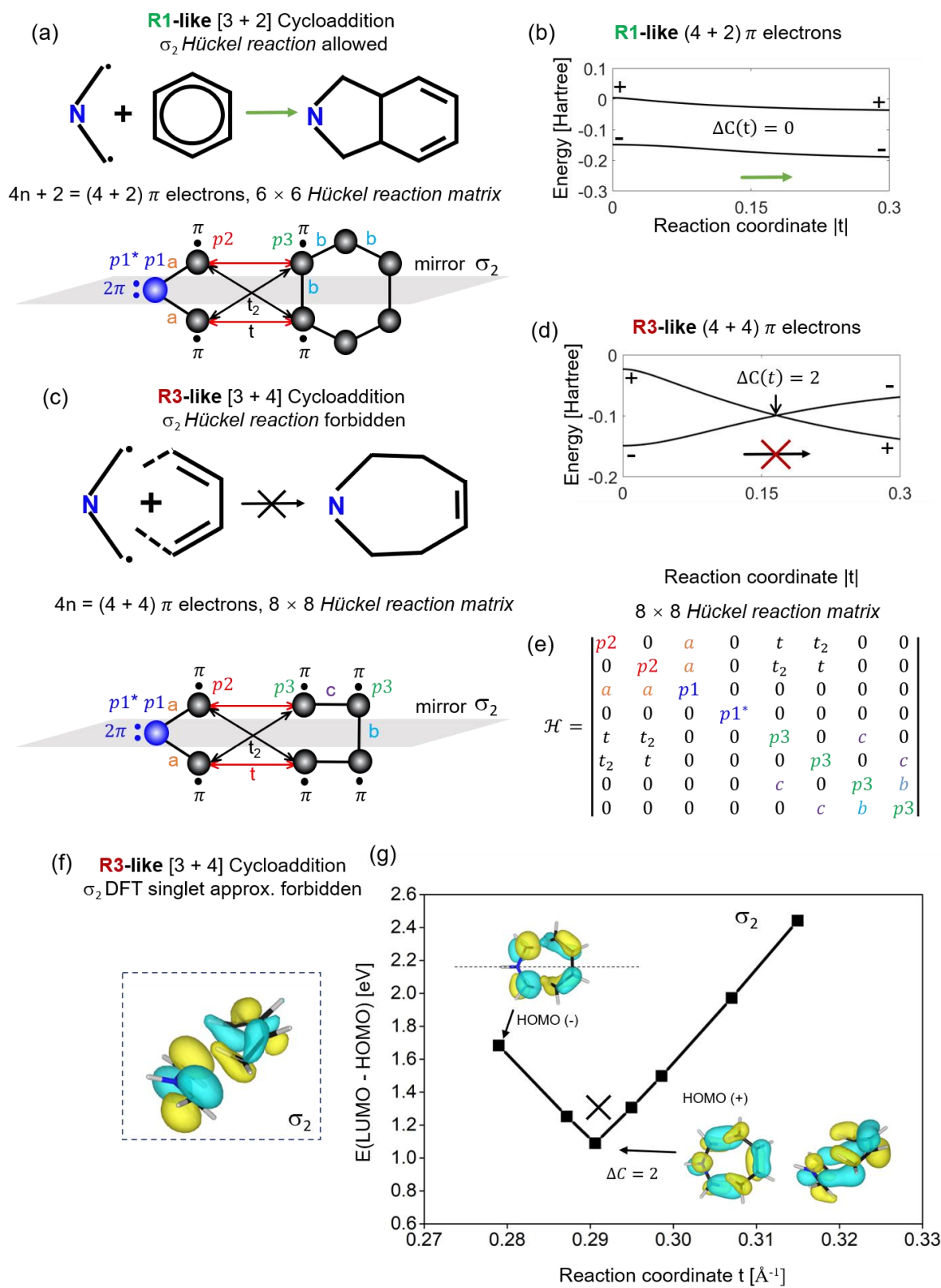

**Fig. 4 | $\sigma_2$–Topological Woodward-Hoffmann classification preserving mirror $C_{2h}$ orbital symmetry for aziridine+benzene *Hückel reaction* and DFT models**. **(a)** The **R1**-like reaction of aziridine diradicaloid as a PAMY model with benzene and the reaction model scheme with π electrons as sites. **(b)** The HOMO and LUMO evolution of (a) as t changes, fitting with DFT results: *a* = -0.138, *b* = -0.125, *p1* = -0.216, *p1\** = 0.065, *p2* = -0.149, *p3* = -0.121, $t_2$ = 0.2t in units of Hartree. The topological invariant per molecular orbital is C(t)=$N_+$(t)-$N_-$(t),

see text and **Fig. S2**. **(c)** The **R3**-like reaction of the diradicaloid with butadiene and the reaction model scheme. **(d)** The HOMO and LUMO evolution of (c) as t changes, fitting with DFT results: $a = -0.138$, $b = -0.088$, $c = -0.141$, $p1 = -0.216$, $p1^* = 0.065$, $p2 = -0.149$, $p3 = -0.127$, $t_2 = 0.2t$ in units of Hartree. The crossing in (c) can be classified topologically, **Figure S1**, **S2** **(e)** *Hückel reaction* matrix description of (c), with t and $t_2$ as variables. Note that we assume the interaction between bonding orbitals ($p2$) and antibonding orbitals ($p1^*$) is very weak (~ 0). **(f, g)** DFT study of the reaction modelling in (c, d) showing an approximate crossing (see text).

## Summary

We studied the on-surface and solid-state reaction of PAMY with pentacene from a Woodward-Hoffmann topological classification point of view, and computationally identified pathway **R1** as most plausible. Pathway **R1** entails a Woodward-Hoffmann topologically allowed, highly endothermic, de-aromatization of pentacene and subsequent dehydrogenation yielding a novel internally N-containing polycyclic aromatic **P1** on Ag(100) and in the solid-state. In addition, we studied a plausible barrierless pathway **R2**, with pre-dehydrogenated pentacene, opening up avenues for the design of more efficient reaction pathways with PAMY. A more reactive pathway **R3**, plausible in the solid-state, is rationalized as Woodward-Hoffmann forbidden from frontier orbital analysis and topologically Woodward-Hoffmann forbidden in a symmetry-enforced tight-binding, *Hückel reaction*, model. Topological Woodward-Hoffmann models offer a pedagogic entry to the analytical study of mathematically-defined reaction pathways and radicaloid cycloadditions, and pave the road for the engineering of solid-state or on-surface AB-type cycloaddition polymerization and related nanographenes; building on three levels of interdisciplinarity common in quantum matter design: Topological classification of analytical models, verification or parameterization with quantum chemistry, and experimental realization.

**Methods**

*Experimental:* The precursor (DBAP molecule) was characterized by NMR-spectroscopy in $d_2$-dichloromethane (SI). The exact molecular weight of DBAP and pentacene additives was detected by high resolution matrix-assisted laser desorption/ionization time of flight (HR-MALDI-TOF) mass spectrometry (MS, see SI, **Fig. S15**).

Samples were prepared in UHV chamber under the base pressure below $5.0 \times 10^{-9}$ mbar. The Ag substrate was cleaned by argon ion sputtering (800 V, 4.5 mA, 15 min) and annealing (flash heating to 450 °C) for several cycles. DBAP and pentacene were deposited on Ag(100) substrate using organic molecular beam epitaxy (OMBE) method from a quartz crucible, which was held at 300 °C and 180 °C for 4 min respectively. Subsequently, the sample was moved to SPM chamber under the base pressure below $2.0 \times 10^{-10}$ mbar. After cooling down to ~ 20K, STM measurements were performed using a commercial low temperature scanning tunneling microscope (LT-STM) with a tungsten tip. It worked at constant current mode with bias at 300 mV and tunneling current at 30 pA.

*Computational:* Broken-symmetry density functional theory (BS-DFT) calculations were performed in gas phase utilizing unrestricted B3LYP functional[79-82] with Grimme's dispersion correction and Becke–Johnson damping [D3(BJ)].[83,84] Full geometry optimization and intrinsic reaction coordinate calculations were initially performed using the minimal basis set[85] and then reoptimized using def2-TZVP basis set[86] in ORCA quantum chemistry package[87]. The Intrinsic Bond Orbital (IBO) analysis is a powerful tool for understanding molecular structure and bonding. It localizes molecular orbitals directly from the wavefunction, enabling the interpretation of the electronic structure in chemical systems[88,89]. The intrinsic bond orbital (IBO) analysis was done in PySCF[90] using the wavefunctions obtained in ORCA for singlet and triplet PAMY. The DFT gaps and the reaction energy profile in **Fig. S17** were obtained using 6-31G(d) basis set[91] in Gaussian 09 Software[92] by varying inter-molecular distance. The inter-molecular distance was varied in the dimer system by shifting both molecules slightly along the perpendicular direction of the molecular planes to avoid the geometry frustration when they are too close.

The slab calculations involving Ag(100) substrate are performed using Vienna Ab-initio Simulation Package[93] (VASP). The Perdew-Burke-Ernzerhof (PBE) parametrization[94] of generalized gradient approximation (GGA) is adopted for exchange correlational functions. We considered *van der Waals* (vdW) corrections via Grimme's D3 method[95]. The energy cutoffs of plane-wave basis are set as 400 eV. A single k-point (Γ) is used for structural optimization and the *k*-mesh of $3 \times 3 \times 1$ was used for the density of states. The convergence criteria for the electronic self-consistent loop and atomic structural optimization are $10^{-6}$ eV for electronic energy and 0.02 eV/Å for the atomic force, respectively. The energy profile for detaching one hydrogen atom is calculated in the climb image nudged elastic band (CI-NEB) scheme.

**Acknowledgments**

This research was financially supported by the EU Graphene Flagship (Graphene Core 3, 881603), the Chinese Academy of Sciences (nos. QYZDBSSW-SLH038, XDB33000000, XDB33030300), National Natural Science Foundation of China (Grant Nos. 11974403, 11974045, 62488201), the German Research Foundation (DFG) within the Clusters of Excellence "Center for Advancing Electronics Dresden (cfaed)", "Matters of Activity", the DFG-SNSF Research Project (EnhanceTopo, No. 429265950), the K. C. Wong Education Foundation of Chinese Academy of Sciences, the National Key Research and Development Program of China (Grant No. 2020YFA0308800) and the Alexander von Humboldt Foundation. We thank Tilo Lübken (Dresden University of Technology) for NMR measurements, Knud Seufert for help with the STM. We gratefully acknowledge Akimistu Narita, Oliver Dumele and Prince Ravat for critical discussions.

**Author contributions**

A. M., W.-H. D. and L. M. performed the theoretical analysis. J. L., W. A., J.V. B. and C.-A. P. performed the microscopy measurements and data interpretation. A. M., L. M., W.-H. D., X.-S. D. and J.-T. S. performed and coordinated the calculations. K. L., M. R., X. W., R. B., J. M., A. N., K. M and X. F. performed MS measurements and synthesized the compounds. C.-A. P. designed the research and supervised the project. All authors participated in discussing and editing the manuscript.

J. L., A. M. and W.-H. D. contributed equally to this work.

**Competing interests**

The authors declare no competing interests.

*Supplementary Information for*

# Topological Woodward-Hoffmann classification for cycloadditions in polycyclic aromatic azomethine ylides

Juan Li[1,#], Amir Mirzanejad[5,#], Wen-Han Dong[2,#], Kun Liu[3], Marcus Richter[3], Xiao-Ye Wang[6,7], Reinhard Berger[3], Shixuan Du[2], Willi Auwärter[4], Johannes V. Barth[4], Ji Ma[3], Klaus Müllen[6], Xinliang Feng[3], Jia-Tao Sun[8], Lukas Muechler[5], Carlos-Andres Palma[2,9]

[1]*Advanced Research Institute for Multidisciplinary Science, Beijing Institute of Technology, 100081 Beijing, China.*

[2]*Institute of Physics & University of Chinese Academy of Sciences, Chinese Academy of Sciences, 100190 Beijing, China.*

[3]*Chair for Molecular Functional Materials, Center for Advancing Electronics Dresden (cfaed), Faculty of Chemistry and Food Chemistry, Dresden University of Technology, Mommsenstr. 4, 01062 Dresden, Germany.*

[4]*Physics Department E20, Technical University of Munich, James-Franck-Str. 1, 85748 Garching, Germany.*

[5]*Department of Chemistry, Penn State University, 108 Chemistry Building, 16802 University Park, United States.*

[6]*Max Planck Institute for Polymer Research, Ackermannweg 10, 55128 Mainz, Germany.*

[7]*State Key Laboratory of Elemento-Organic Chemistry, College of Chemistry, Nankai University, 300071 Tianjin, China.*

[8]*School of Integrated Circuits and Electronics, MIIT Key Laboratory for Low-Dimensional Quantum Structure and Devices, Beijing Institute of Technology, 100081 Beijing, China.*

[9]*Department of Physics & IRIS Adlershof - Humboldt-Universität zu Berlin, 12489, Berlin, Germany*

**List of Abbreviations**

| |
|---|
| antisymmetric (AS), symmetric (S) molecular orbital (MO) |
| broken-symmetry density functional theory (BS-DFT) |
| climb image nudged elastic band (CI-NEB) |
| density functional theory (DFT) |
| diradicaloid PAMY (rPAMY) |
| dibenzo-*9a*-azaphenalene (DBAP) |
| *Hückel reaction* model |
| high resolution matrix-assisted laser desorption/ionization time of flight (HR-MALDI-TOF) |
| intermediate (Int1), product 1 (P1), dehydrogenated intermediate (DH) |
| intrinsic bond orbital (IBO) |
| intrinsic reaction coordinate (IRC) |
| matrix-assisted laser desorption-ionization mass spectrometry (MALDI-MS) |
| mass spectrometry (MS) |
| nudge elastic band (NEB) |
| organic molecular beam epitaxy (OMBE) |
| polycyclic aromatic hydrocarbons (PAHs) |
| polycyclic aromatic azomethine ylide (PAMY) |
| scanning tunneling microscopy (STM) |
| topological Woodward-Hoffmann classification |
| transition state (TS) |
| ultra-high vacuum (UHV) |
| Woodward and Hoffmann（WH） |
| 8*H*-isoquinolino[4,3,2-*de*]phenanthridin-9-ium tetrafluoroborate (DBAP salt) |

**List of Figures**

**Reaction Model Discussions**

We explain the reaction matrix model scheme in **Fig. S1**, which we assume the pristine subspaces are unchanged during reaction. Also, it is the start point of our DFT calculations that varying inter-molecular distance is analogous to varying t of the model. Due to the robustness of topological protection, using a simple mirror-preserved path gives consistent result with other paths with slight perturbation.

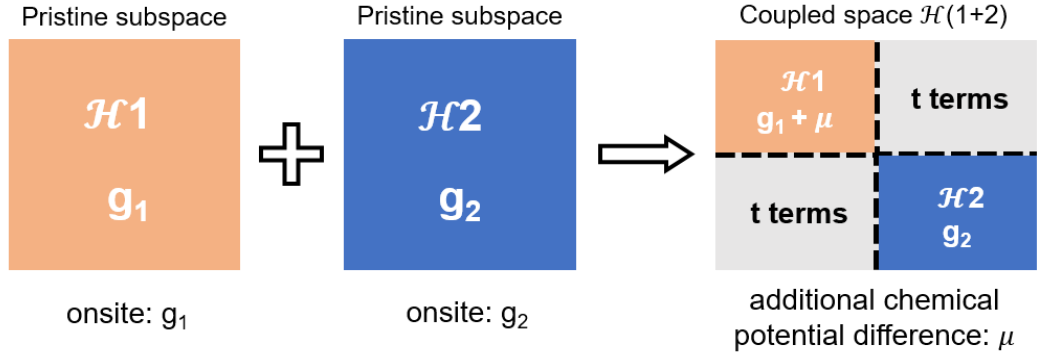

**Figure S1.** Schematics of the *Hückel reaction* model frame. The variable t terms accounts for the process of reaction and additional chemical potential difference $\mu$ comes from different reaction conditions, such as substrates, heating, etc.

In **Fig. S2a** we review the Hamiltonian H for the reaction of the "open aziridine" biradical with butadiene and give an explicit representation J of the mirror symmetry in the basis of the atomic orbitals. As J and H commute, one can block diagonalize H into a J-even (positive J mirror parity) and J-odd (negative mirror parity) subspace displayed in **Fig. S2b**. As there are no matrix elements between the blocks, crossings of eigenstates with different mirror parities cannot be removed unless the mirror symmetry is explicitly broken during the reaction. In **Fig. S2c** we show that there is such a crossing between occupied and unoccupied states of different mirror parity as one tunes t. In order to highlight the topological nature of this crossing[96], we define a topological mirror winding number $N_\pm$ via the single particle Green's function[2] $G^\pm(\omega) = (\omega - H^\pm)^{-1}$ for each block $H^\pm$ of the Hamiltonian with $\omega \in \mathbb{C}$. For a single particle Hamiltonian such as ours, the Greens function is a diagonal matrix with entries $\left(\omega - \epsilon_i^\pm\right)^{-1}$, where $\epsilon_i^\pm$ is the i-th eigenvalue of $H^\pm$, i.e. the eigenvalues of $H^\pm$ are poles of $G^\pm(\omega)$ on the real axis. The winding number is defined along a closed contour $C$ in the complex plane as

$$N_\pm = Tr \frac{1}{2\pi i} \oint_C \frac{1}{G^\pm(\omega)} \partial_\omega G^\pm(\omega)\, d\omega \;=\; P_C \in \mathbb{Z},$$

where $P_C$ is the number of poles enclosed in the contour through the argument principle. Choosing the contour as displayed in **Fig. S2d,** the winding number measures the winding of the diagonal entries of $G^{\pm}(i\omega)$ around 0 in the complex plane and is equal to the number of occupied molecular orbitals of $H^{\pm}$. The winding number is a $\mathbb{Z}$ valued invariant that classifies the mapping $S^1 \to GL(N, \mathbb{C})$ from the circle to the set of complex invertible matrices $GL(N, \mathbb{C})$. We define the mirror topological invariant for each value of the reaction coordinate t as $N(t) = N_+(t) - N_-(t) \in \mathbb{Z}$. The winding number is not defined at the crossing point around $t_c \sim 0.16$, allowing the invariant to jump discontinuously. In case of no crossings between the occupied and unoccupied states with different mirror eigenvalues, the invariant stays constant during the reaction.

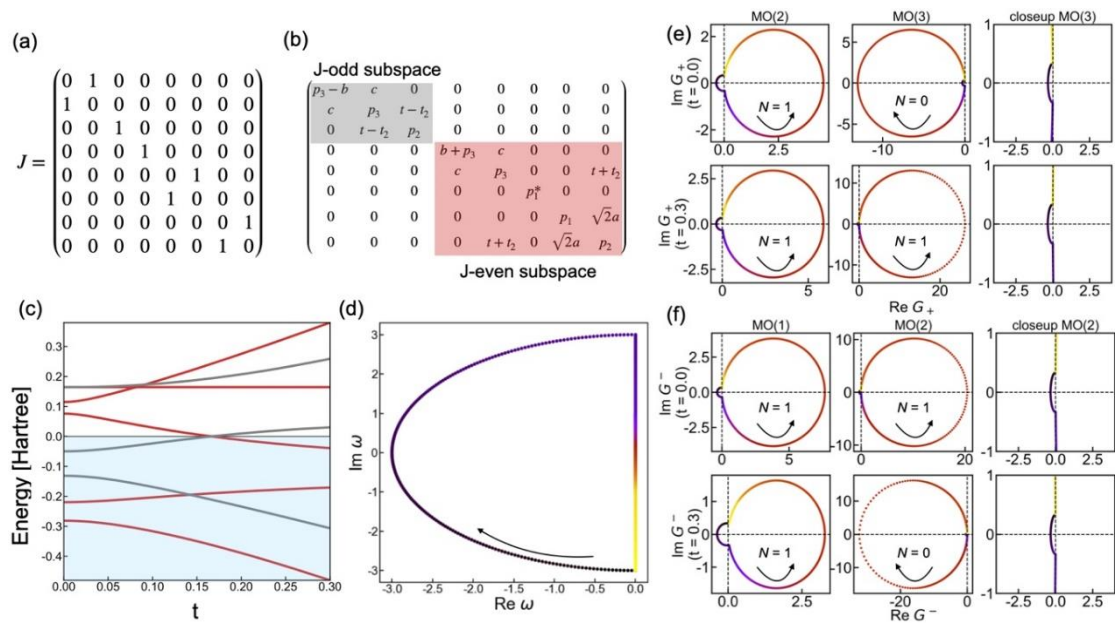

**Figure S2.** The details of topological analysis. **(a)** J matrix (representation of the mirror symmetry in the orbital basis) for the forbidden reaction of the "open aziridine" diradicaloid with butadiene. **(b)** Block-diagonal Hamiltonian in the J-basis. The grey block corresponds to the J-odd while the red block corresponds to the J-even part of the Hamiltonian. **(c)** Energy levels as a function of t for calculated by diagonalizing the block-diagonal Hamiltonian using $a = -0.138$, $b = -0.088$, $c = -0.141$, $p1 = -0.216$, $p1^* = 0.065$, $p2 = -0.149$, $p3 = -0.127$, $t_2 = 0.2t$ in units of Hartree. A chemical potential has been added to move the crossing point to 0 energy. J-even states are colored red, while J-odd states are colored grey. The occupied states are highlighted in light blue. **(d)** Contour used to evaluate the winding number. The arrow indicates the orientation of the contour and black colors indicate the starting point **(e)** Winding number of the 2nd and 3rd molecular orbital of $H^+$ at t = 0 and t = 0.3 for the contour displayed in (d). The 3rd MO winds counterclockwise around 0 with a winding number of 1 for t = 0 and 0.3. The 3rd MO winds around 0 for t = 0.0 with a winding number of 1, but does not wind around 0 for t = 0.3 as shown in the inset. **(f)** Winding number of the 2nd and 3rd molecular orbital of $H^-$ at t = 0 and t = 0.3 for the contour displayed in (d). The 1st MO winds counterclockwise around 0 with a winding number of 1 for t = 0 and 0.3. The 2nd MO winds around 0 for t = 0.0 with a winding number of 1 but does not wind around 0 for t = 0.3 as shown in the inset.

## DFT Based Parameterization

In **Fig. S3**, we show the significance of the choice of main action sites used because it determines which term is the dominant t and leads to different products. The pristine subspace of ethylene and *cis*-butadiene molecule are achieved by DFT-fitted parameters. It is found that the contribution of minor $t_2$ terms should not be ignored since it embodies the competition of active sites and less active sites during the cycloaddition.

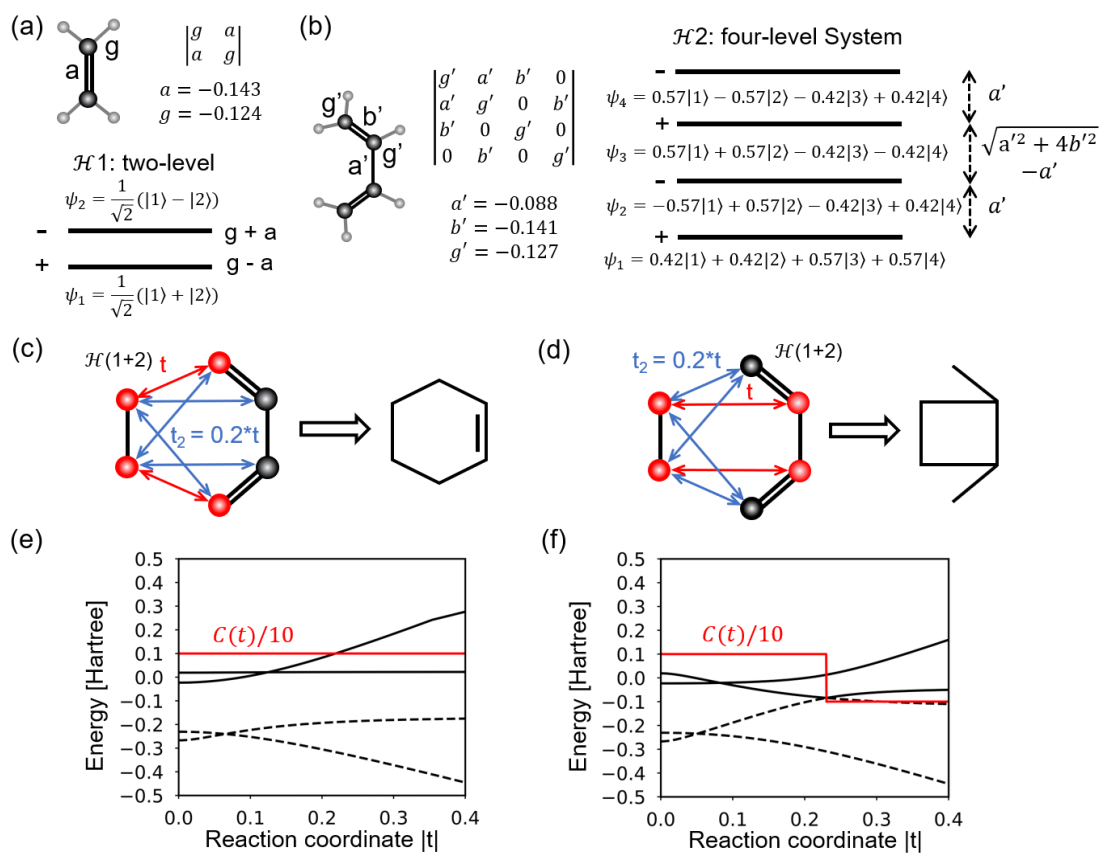

**Figure S3.** DFT fitted ethylene + *cis*-butadiene reaction. **(a)** Hückel-like matrix description with DFT fitted parameters (in units of Hartree) of ethylene. **(b)** Hückel-like matrix description with DFT fitted parameters of *cis*-butadiene. In (a) and (b), + (-) denote mirror parity. **(c, d)** Two possible reaction paths and products (without hydrogens). $t_2$ terms are chosen as $t_2 = 0.2t$ to describe minor but competing reaction possibilities. **(e)** Evolution of MO eigenvalues based on reaction scheme in (c). **(f)** Evolution of MO eigenvalues based on reaction scheme in (d). The red lines in (e) and (f) represent topological invariant $C(t)$ with shrinking ten times. The matrix is $6 \times 6$, where two occupied MOs (dashed) and two unoccupied MOs (solid) are displayed.

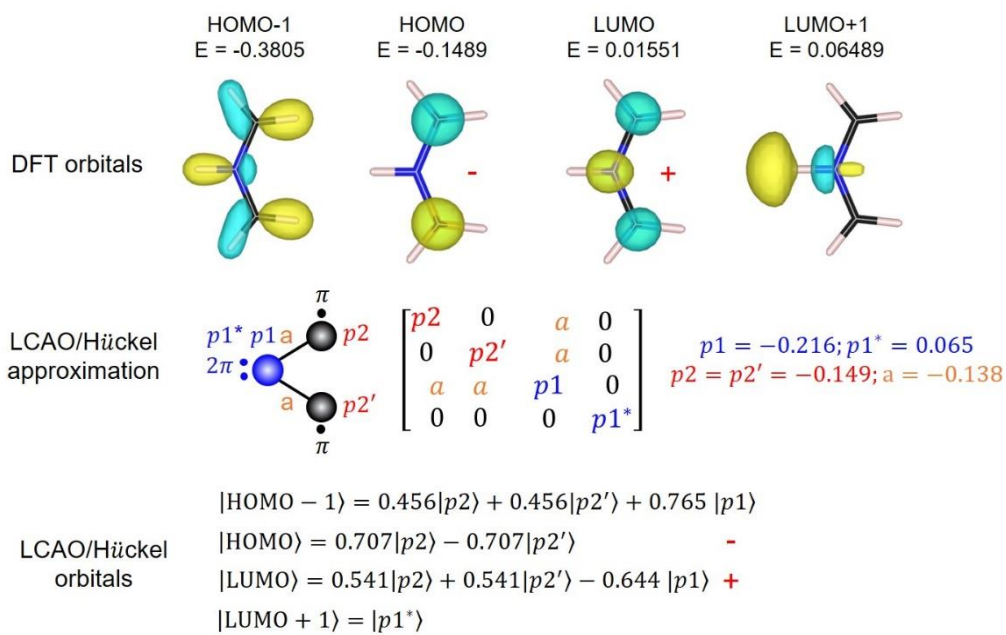

**Figure S4.** Rationality of the Hückel-like parameterization: using aziridine diradicaloid as an example. The iso-surfaces of DFT orbitals are chosen as 0.1 $e/bohr^3$. One can see HOMO and LUMO orbitals are well fitted by our parameterization (including the mirror parities).

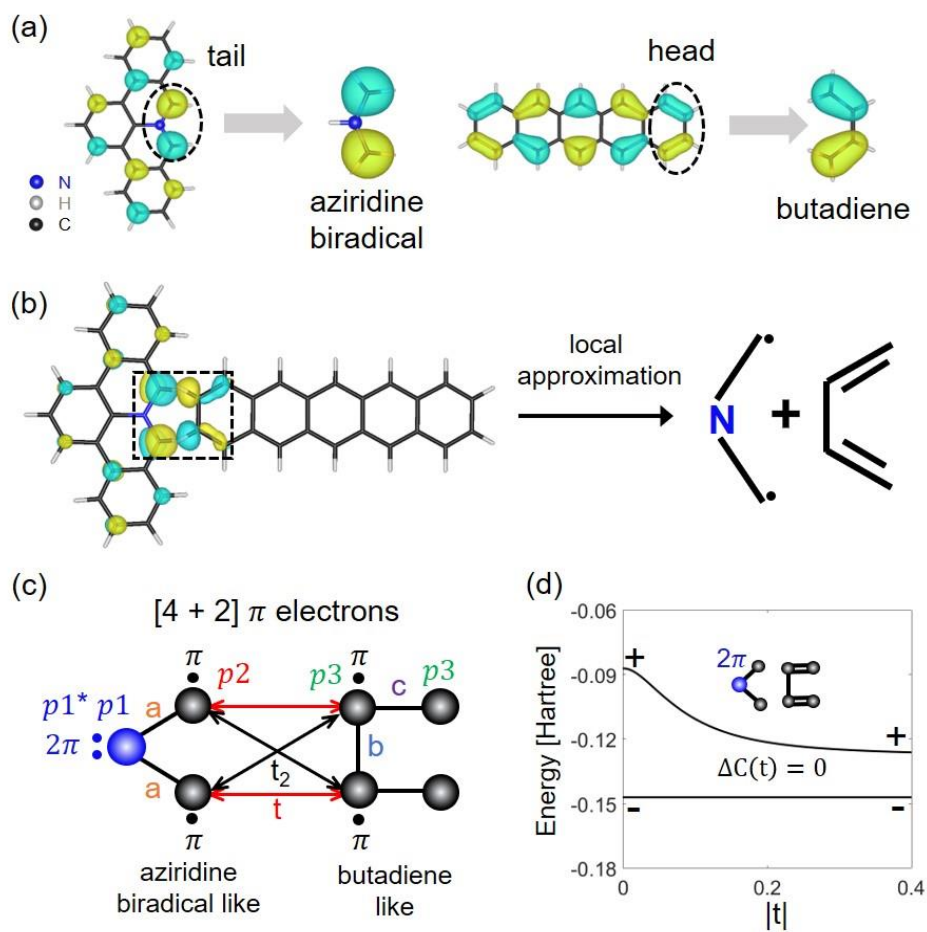

**Figure S5.** Reaction matrix of rPAMY + pentacene reaction under local approximation. **(a)** Orbital resemblance of HOMO of rPAMY and pentacene, with an "open aziridine" diradicaloid and butadiene, respectively. The orbital parity is revealed by blue (yellow), and the iso-surfaces are chosen as 0.04 $e/bohr^3$. **(b)** Local approximation of rPAMY + pentacene reaction due to charge redistribution when approaching. The left panel shows the HOMO when rPAMY and pentacene are sufficiently close. **(c)** Reaction model scheme under local approximation. **(d)** HOMO and LUMO evolution of (c) as t changes, with parameters fitting from DFT results: $a = -0.060$, $b = -0.053$, $c = -0.062$, $p1 = -0.126$, $p1^* = -0.040$, $p2 = -0.147$, $p3 = -0.128$, $t_2 = 0.2t$ in units of Hartree. Here, only the R1 case of **Fig. 3** is considered because the situations of R2 and R3 go beyond Hückel-like parameterization.

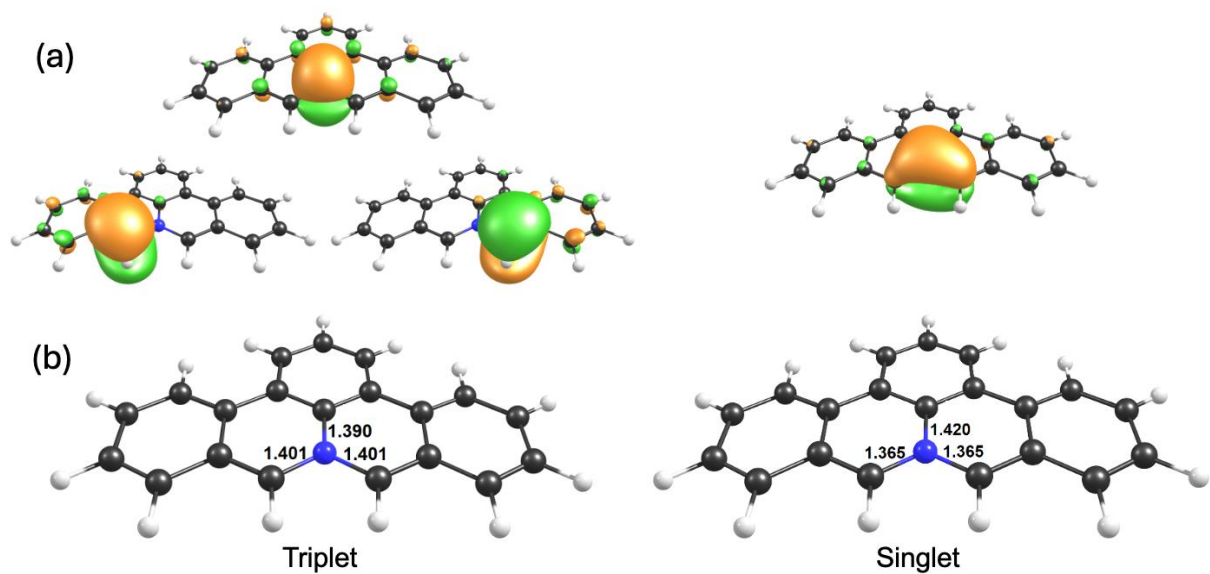

**Figure S6.** IBOs and chosen bond distances of PAMY in both singlet and triplet. **(a)** Intrinsic Bond Orbitals (IBOs) in the fully optimized triplet and singlet PAMY at BS-UB3LYP/def2-TZVP level of theory. **(b)** The C-N interatomic distances (in Å) in the optimized triplet and singlet PAMY.

# Calculations on Ag(100) Surface

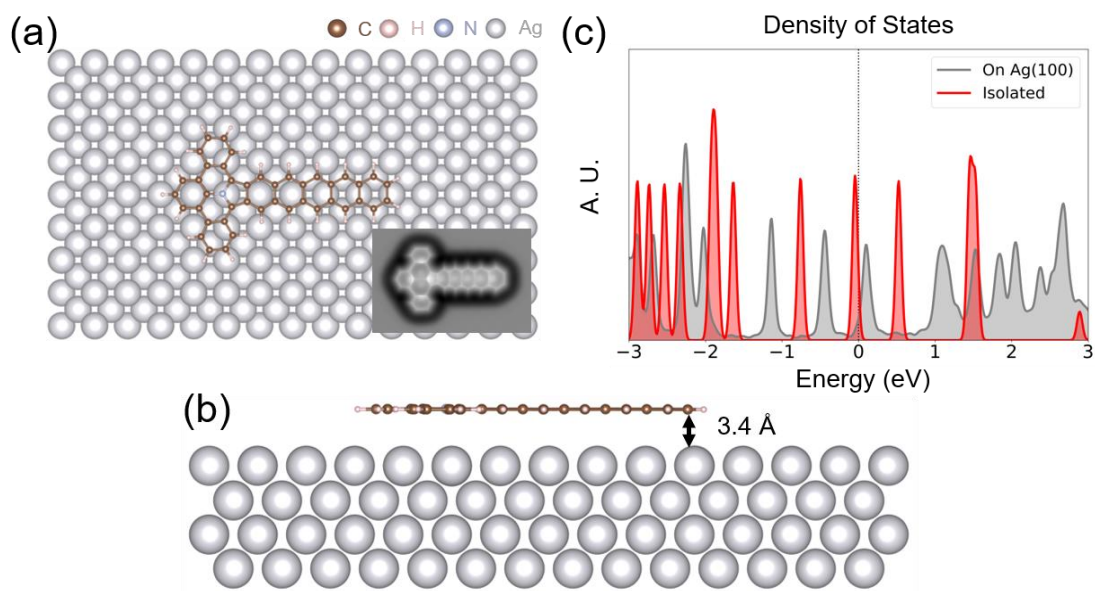

**Figure S7.** Adsorption of PAHs with tetraisoindole core on Ag(100) surface. **(a, b)** Top and side views of the structure. The right bottom panel of (a) is the q-plus AFM simulation. The adsorption energy for one molecule is 4.019 eV. **(c)** Density of states of PAHs with tetraisoindole core under PBE + vdW-D3 level.

**Fig. S8** indirect evidence that dehydrogenation may happen before or after cycloaddition on surfaces since energy barrier ~ 2 eV is accessible between 200 and 400 K[32].

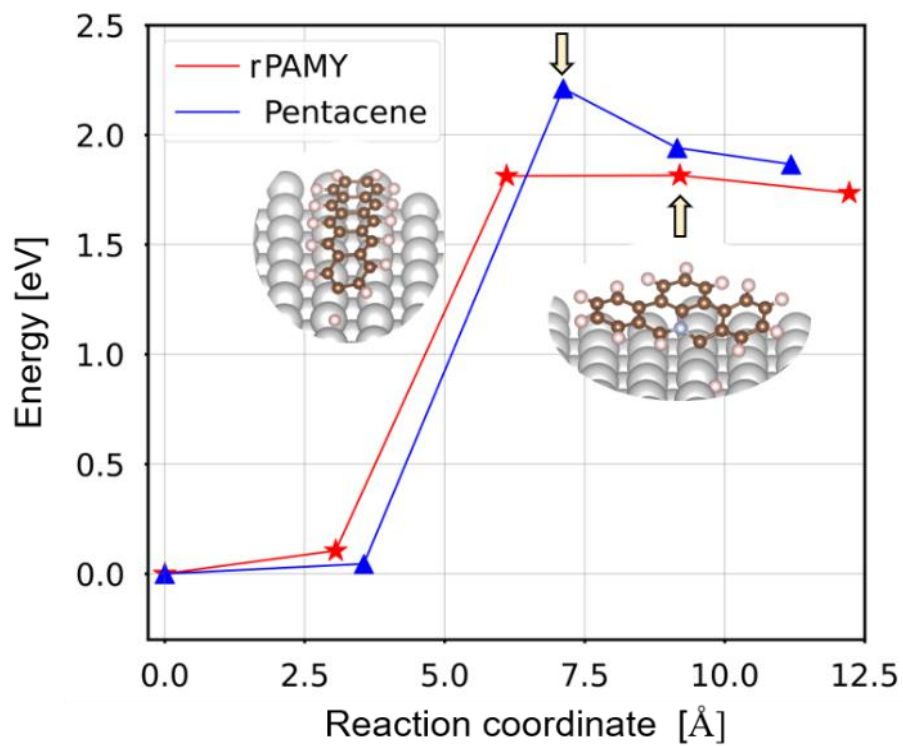

**Figure S8.** Energy profile for rPAMY (pentacene) to detach a hydrogen on Ag(100) surface. The energy barrier is 1.82 eV (2.21 eV) for rPAMY (pentacene). The insets show corresponding transition state structures.

**Calculated Molecular Properties**

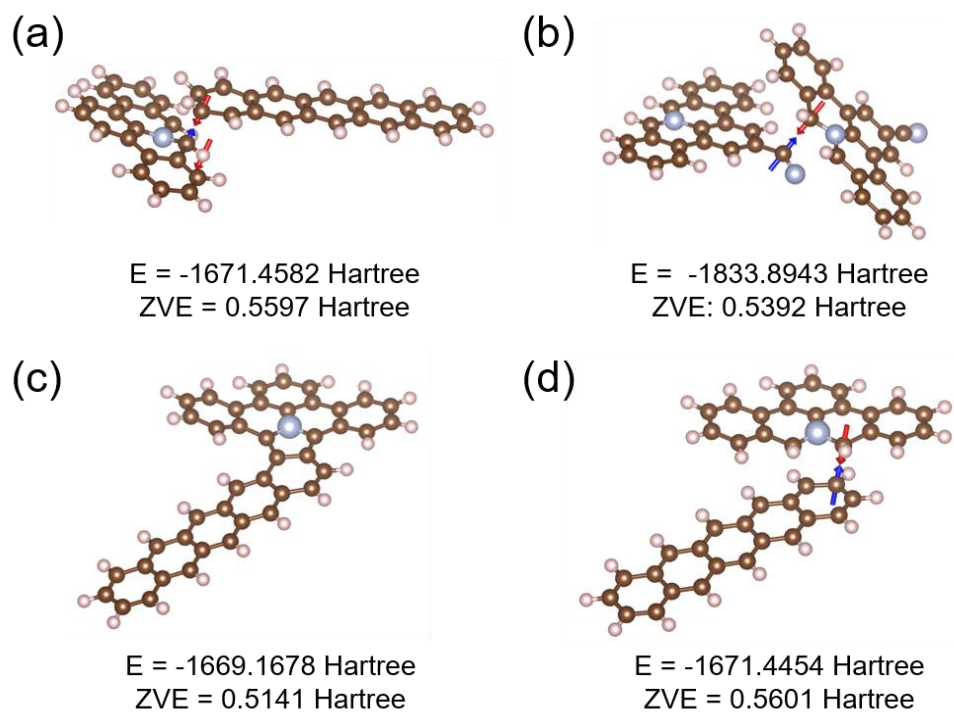

**Figure S9.** Calculated transition states. **(a)** Transition state in **Fig. 2e**. ZVE is short for zero-point vibrational energy. The arrows denote the only virtual vibrational modes. **(b)** Transition state in **Fig S17f**. **(c)** Side product P2. **(d)** Transition state before forming the side product P2 in (c). The total energy of (d) is 0.36 eV higher than that of (a), deciding the side product is lower in yield (~ 20%).

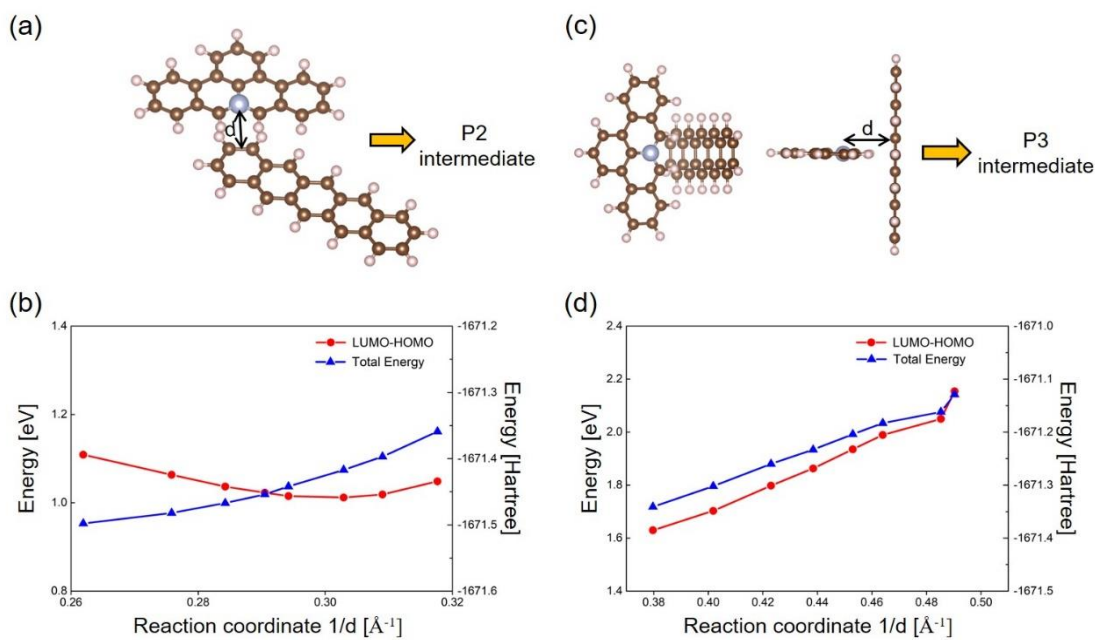

**Figure S10.** DFT determination of the reactivity of P2 and P3. **(a)** Schematic of the chosen reaction path in forming the intermediate of P2 and **(b)** corresponding energy evolution along the reaction path. **(c)** Schematic of the mirror symmetric reaction path in forming the intermediate of P3 and **(d)** corresponding energy evolution along the reaction path. The DFT results indicate both P2 and P3 are obtainable because there is no singularity of LUMO-HOMO (or total energy), while P3 may not be observed due to the geometrical obstruction during surface reaction and the total energy in (d) is around 0.2 Hartree higher than that of (b) when approaching the intermediate states.

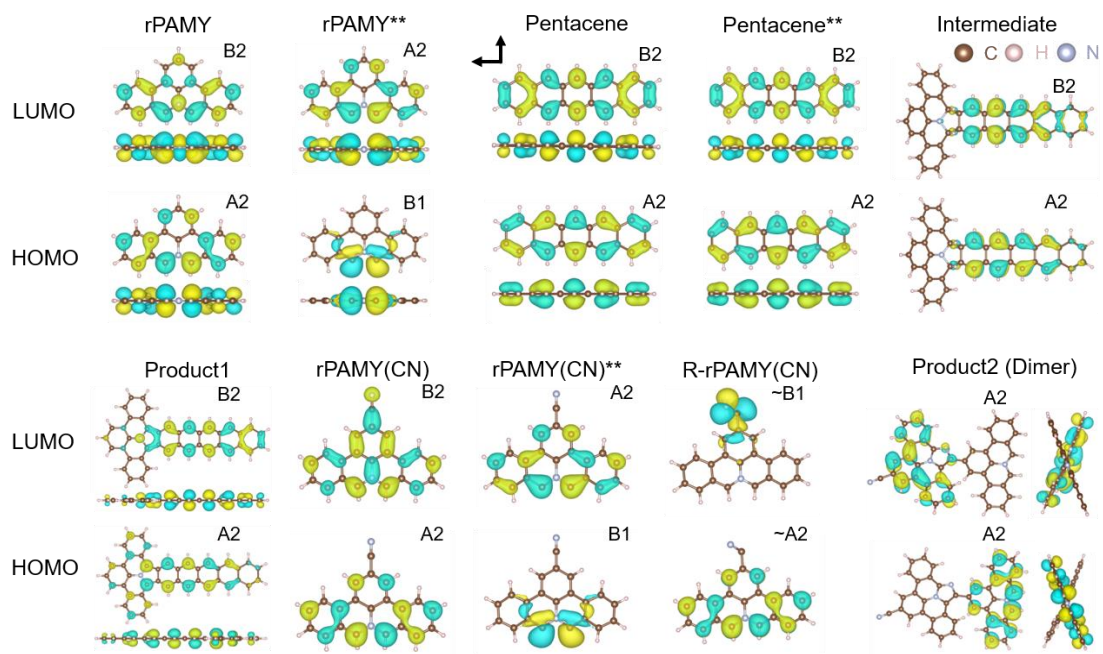

**Figure S11.** Symmetry of the frontier orbitals. The iso-surfaces are 0.03 $e/bohr^3$.

## General Experimental Methods

All the reagents were obtained from Sigma Aldrich, TCI or Strem. All these chemicals were used as received without further purification. Solvents employed for work-up and column chromatography were purchased in HPLC quality and used directly without further purification. NMR spectra were recorded on a Bruker AV-II 300 spectrometer operating at 300 MHz for $^1$H and at 75 MHz for $^{13}$C at room temperature. All chemical shifts are reported in parts per million (ppm). High-resolution mass spectrometry (HR-MS) was performed on a Bruker Reflex II-TOF spectrometer using a 337 nm nitrogen laser by matrix assisted laser desorption/ionization (MALDI) with *trans*-2-[3-(4-*tert*-butylphenyl)-2-methyl-2-propenylidene]malononitrile (DCTB) as the matrix. (2'-Amino-[1,1':3',1''-terphenyl]-2,2''-diyl) dimethanol (**S1**) and 8*H*-isoquinolino[4,3,2-*de*]phenanthridin-9-ium tetrafluoroborate (**1**) were synthesized according to our previous report[28].

## Precursor Synthesis

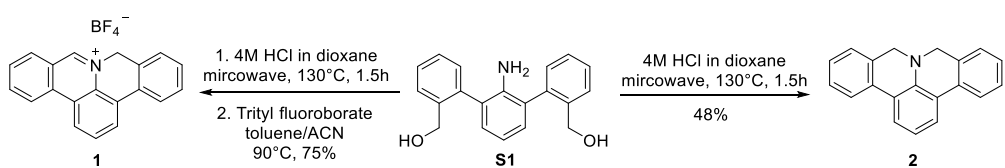

**Scheme S1:** Synthetic scheme toward 8*H*-isoquinolino[4,3,2-*de*]phenanthridin-9-ium tetrafluoroborate (DBAP salt, **1**) and dibenzo-*9a*-azaphenalene (DBAP, **2**) starting from (2'-amino-[1,1':3',1''-terphenyl]-2,2''-diyl)dimethanol (**S1**).

## Synthesis of dibenzo-*9a*-azaphenalene (DBAP, 2)

In a dry and inert microwave tube, compound **S1** (3.43 mmol, 1.00 eq) was added in anhydrous hydrogen chloride solution (4.0 M in dioxane, 10 mL). The tube was then capped and placed in a microwave reactor. A dynamic mode was chosen (300 W, power max: on, activated cooling, pre-stirring: 10 s, temperature: 130 °C) for 90 min. After cooling of the reaction mixture to room temperature, the reaction tube was transferred to the freezer of the glovebox for crystallization process. After overnight, the precipitates were filterated and washed with MeOH. The title compound **2** was observed as green solid in a yield of 48%.

**$^1$H NMR (300 MHz, CD$_2$Cl$_2$):** δ: 7.69-7.67 (m, 4H,), 7.35 (td, J = 7.6 Hz, J = 7.6 Hz, J = 1.3 Hz, 2H), 7.28-7.25 (m, 2H), 7.21-7.20 (m, 2H), 6.96-6.94 (m, 1H), 4.25 (s,4H) ppm.

**$^{13}$C NMR (75 MHz, CD$_2$Cl$_2$):** δ: 142.71, 130.89, 130.20, 127.27, 126.69, 125.23, 122.49, 121.65, 121.59, 118.86, 53.21 ppm.

**HR-MS (MALDI-TOF):** m/z ([M-H]$^+$) = 268.112 [M-H]$^+$, calcd. for C$_{20}$H$_{15}$N: m/z = 269.120.

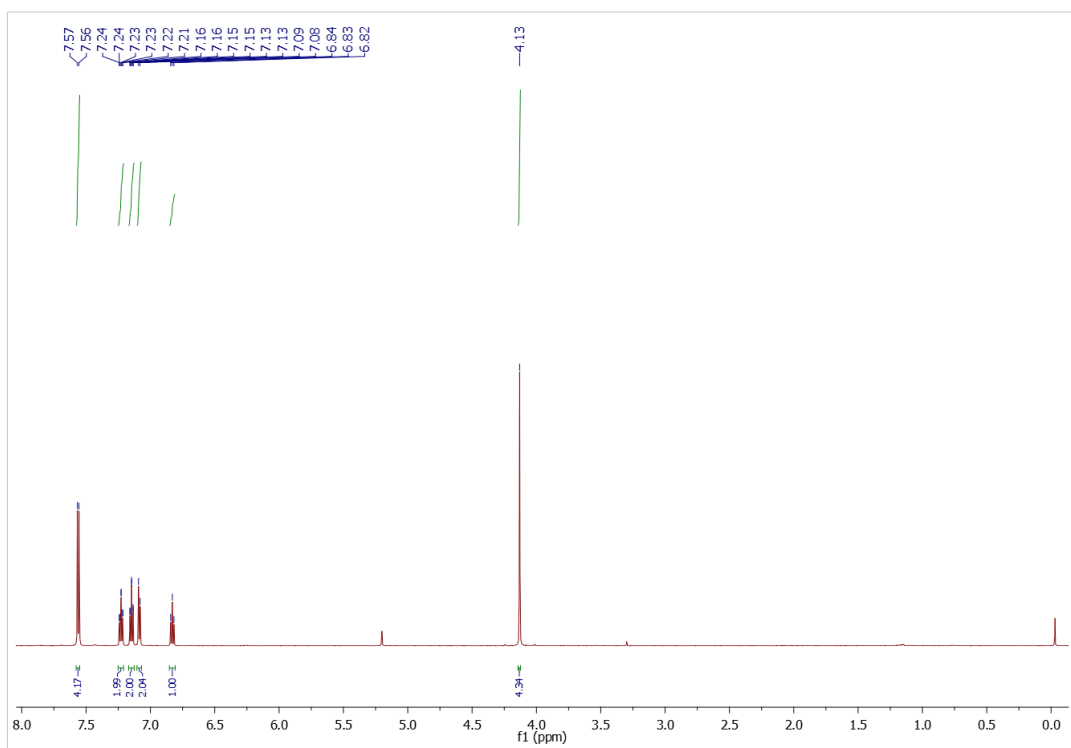

**Figure S12.** $^1$H NMR spectrum of **2** (solvent: CD$_2$Cl$_2$).

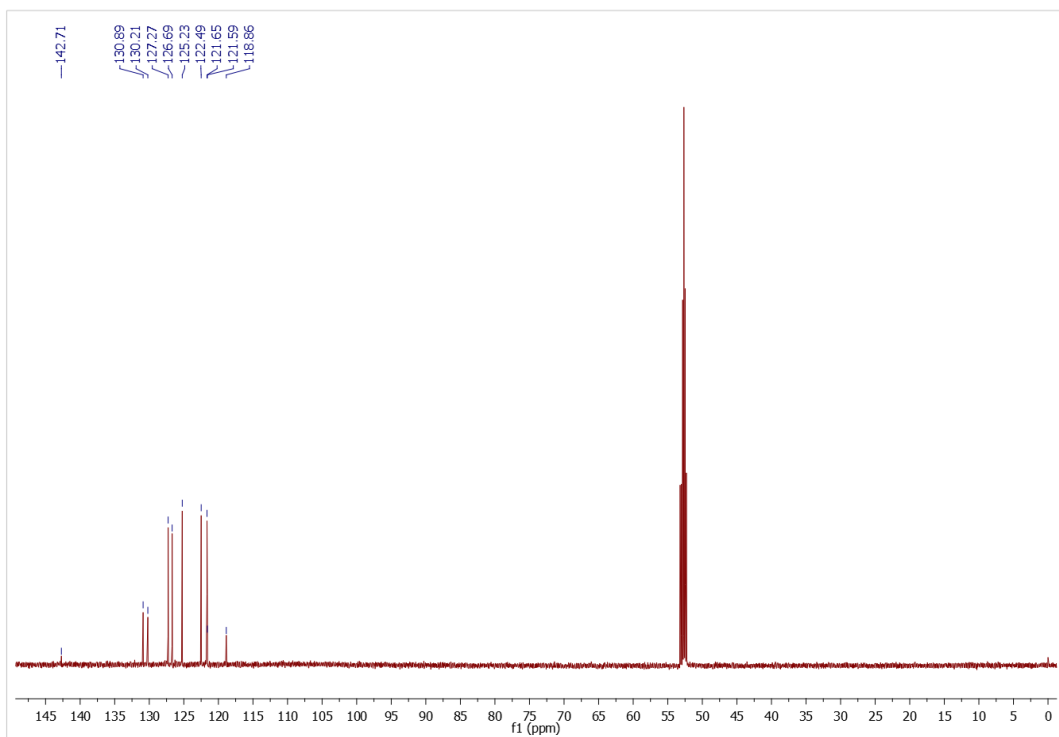

**Figure S13.** $^{13}$C NMR spectrum of **2** (solvent: $CD_2Cl_2$).

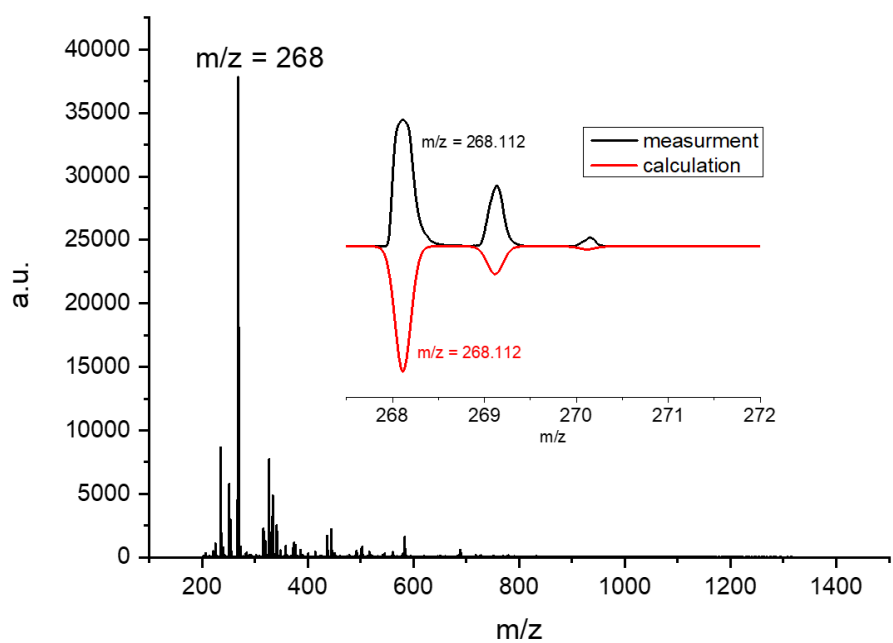

**Figure S14.** MALDI-TOF mass spectrum of compound **2**.

**Procedure for the Solid-State Synthesis and MALDI-TOF MS Analysis**

We demonstrated that the thermally induced reaction between the rPAMY precursor **1** and pentacene occurs in the solid state. In details, compound **1** and pentacene (molar ratio, 1:1) were pressed into a pill form (Specac compression device, seven tons) and put into glass ampoule. The ampoule was sealed under vacuum after three-pump-thaw cycles (~ $10^{-2}$ mbar) and then heated to 250 °C for 72 h. After cooling to room temperature, the crude mixture was analyzed by MALDI-TOF mass spectrometry without further purification (**Fig. S15a**). As shown in **Fig. S15b**, the mass peak of the dihydro-intermediate of [rPAMY + pentacene] (**iv**) (m/z = 543.198) and the mixtures of the dimerization of **1** (hexabenzo[*b,b',g,g',ij,i'j'*]pyrazino[2,1,6-*de*:3,4,5-*d'e'*]diquinolizine (**i**), its dihyro- and tetrahydro-precursors (**ii** and **iii**) were detected. Interestingly, we also observed the byproduct (**v**) of [2×rPAMY + pentacene] (m/z=806.272) (**Fig. S15c**). In order to get the fully dehydrogenated products, we added the oxidant 2,3-dichloro-5,6-dicyano-1,4-benzoquinone (DDQ) in the reaction mixture (**1**, pentacene and DDQ (molar ratio, 1:1:2.2)) following the same solid synthesis procedure. As expected, the MALDI-TOF mass spectra show the successful dehydrogenated products (**Fig. S15d-f**), including the hexabenzo[*b,b',g,g',ij,i'j'*]pyrazino[2,1,6-*de*:3,4,5-*d'e'*]diquinolizine (**i**), the dehydrogenated products of [rPAMY + pentacene] (**vi**) and the dehydrogenated products of [2 × rPAMY + pentacene] (**vii**).

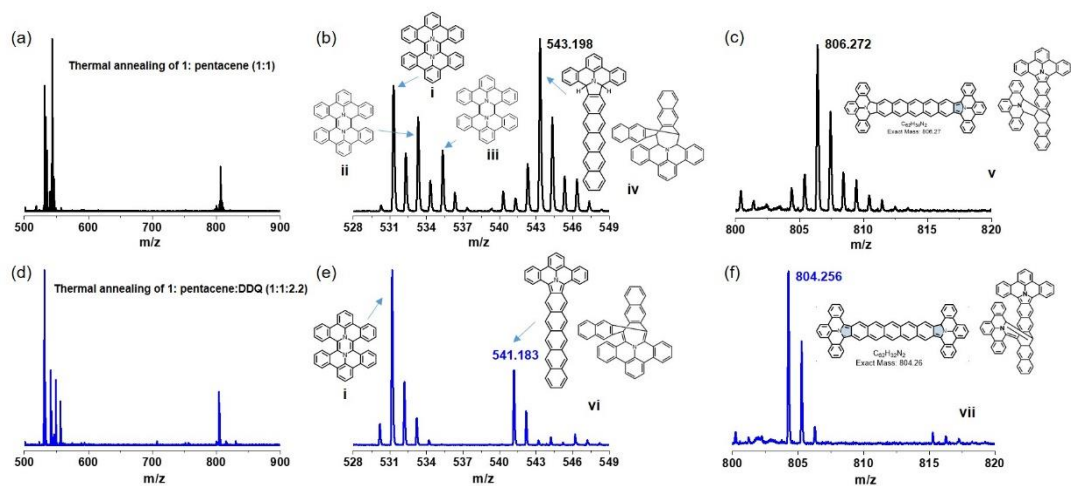

**Figure S15.** Solid-state synthesis of thermal annealing of **1** and pentacene. **(a)** The MALDI-TOF mass spectrum after thermal annealing of **1** and pentacene (molar ratio, 1:1). **(b)** and **(c)** are the magnified regions in (a). **(d)** The MALDI-TOF mass spectrum after thermal annealing of **1**, pentacene and DDQ (molar ratio, 1:1:2.2). **(e)** and **(f)** are the magnified regions in (d). Tentative assignments shown as insets.

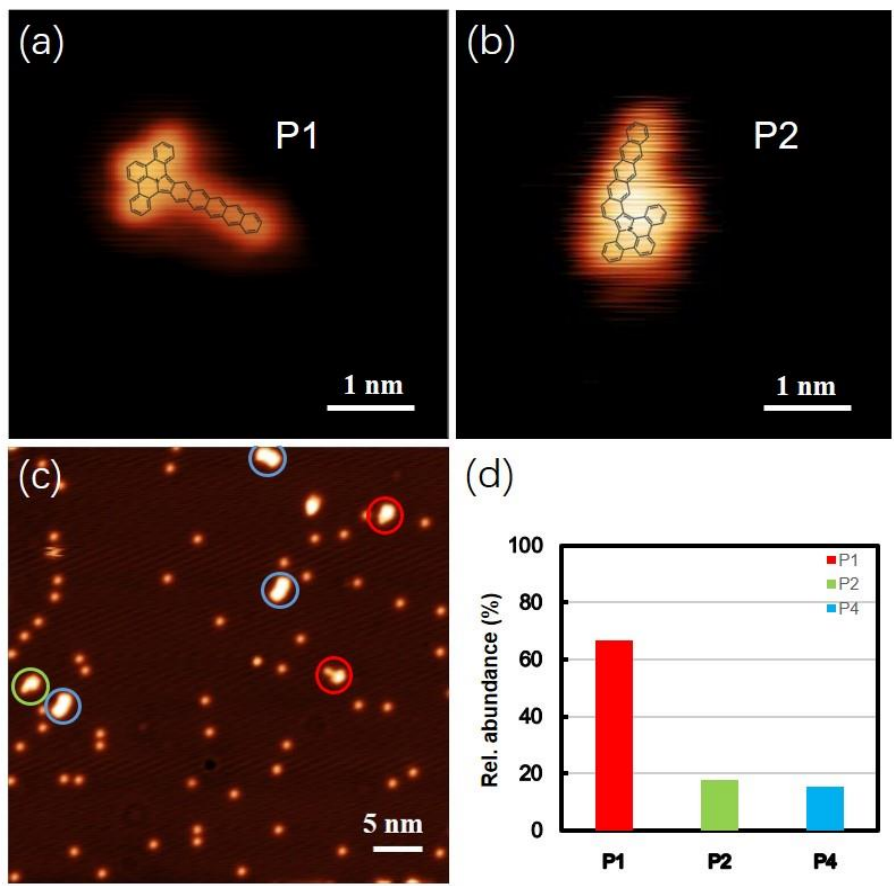

**Figure S16.** STM of different products. **(a)** Zoom-in high-resolution STM of the reaction main product (P1) on Ag(100) after on-surface synthesis at ~ 400 °C. $I_t$ = 30 pA, $V_s$ = 300 mV. Scale bar: 1 nm. **(b)** Zoom-in STM of the minor non-symmetrical side product (P2) on Ag(100) after on-surface synthesis at ~ 400 °C. $I_t$ = 30 pA, $V_s$ = 300 mV. Scale bar: 1 nm. **(c)** STM of the reaction compounds on Ag(100) after on-surface synthesis at ~ 400 °C. $I_t$ = 30 pA, $V_s$ = 300 mV. Main product (P1), side product (P2) and two rPAMY + pentacene product (P4) are circled in red, green and blue, respectively. Scale bar: 5 nm. **(d)** Relative abundances for reaction product P1, side product P2 and P4.

## Cycloaddition Between Cyano-rPAMY

To generalize our observations to the polymerization of cyano-rPAMY, we have further explored the reaction with an approximate (mirror) symmetry pathway by DFT. Notably, the cyano-head rotation, which can be induced by thermal annealing, results in a marked LUMO charge redistribution (**Fig. S1b**) that activates the reactant. Two possible reaction pathways R1 and R2 consider dehydrogenation order (**Fig. S17c**). A reaction pathway R2 involving the aziridine heterocycle is not allowed in agreement with the previous observations (**Fig. S17e**), and the dehydrogenation barrier is much higher than the transition state from the coordinated rotation of two molecules (**Fig. S17f**). The potential topological singularity of R2 manifests itself in the change of orbital symmetries during the hypothetical reaction process as t varies. Though the symmetry assumed here is approximate, the consistency of DFT results with our previous experiment[72] evidences that our method can be extended to highly-symmetric reaction pathways to uncover the detailed reaction mechanism.

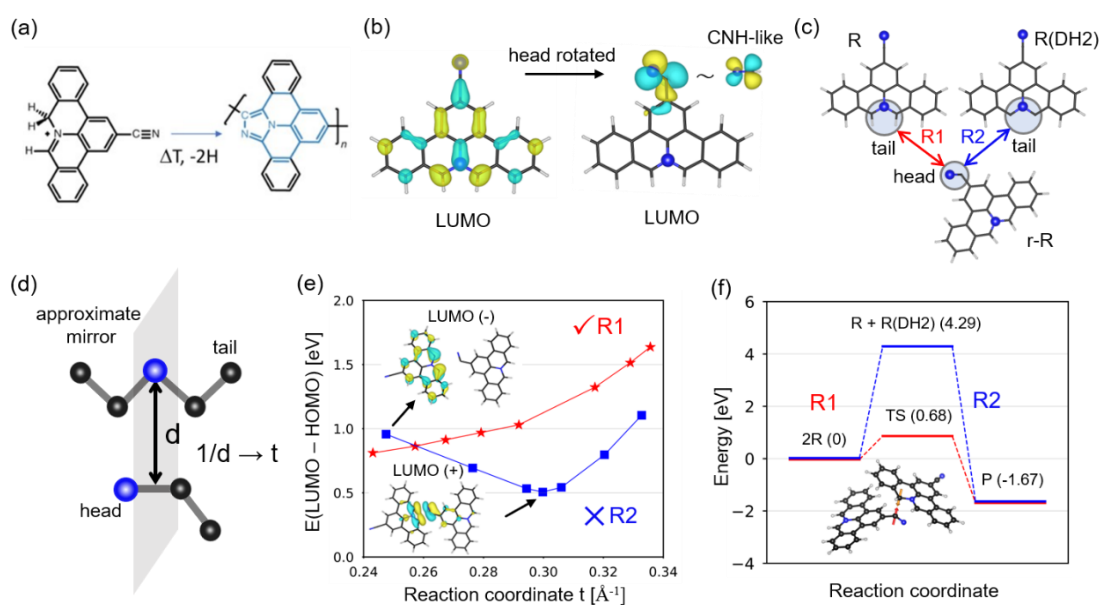

**Figure S17.** The cyano-rPAMY dimer case with approximate mirror symmetry. (**a**) Experimentally reported homocoupling reaction of cyano-rPAMY precursor cyano-DBAP[72] on insulating layers and in the solid state. (**b**) LUMO charge distribution of cyano-rPAMY vs. head-rotated cyano-rPAMY. The rotation entails a charge redistribution, locally assembling the hydrogen cyanide (CNH). Isovalues are 0.04 $e$/bohr$^3$. (**c**) Possible evolution of cyano-rPAMY with reaction paths R1 and R2, where R stands for the reactant, r-R is the rotated reactant and DH2 is two folds hydrogen abstraction. (**d**) Schematic of the approximate mirror symmetry. (**e**) DFT calculated HOMO/LUMO gap evolution of the dimerization in two reaction paths. The insets give the LUMO distributions of R2, where the singularity is topologically obstructed, as indicated by the altered orbital symmetry. (**f**) Energy diagram illustrating R1 and R2, where TS stands for transition state.

## Supplementary Tables

|  | **LUMO+1** | **LUMO** | **HOMO** | **HOMO-1** | **E(singlet)** | **ZVE** |
|---|---|---|---|---|---|---|
| rPAMY | -0.03962(+) | -0.05105(+) | -0.14690(-) | -0.22136(-) | -824.7166 | 0.270253 |
| rPAMY** | -0.04827(-) | -0.09444(-) | -0.17131(-) | -0.19156(+) | -823.3701 | 0.245214 |
| Pentacene | -0.03450(+) | -0.08781(+) | -0.16888(-) | -0.21610(-) | -846.7998 | 0.287526 |
| Pentacene** | -0.07403(-) | -0.09216(+) | -0.17234(-) | -0.22039(-) | -845.4581 | 0.262038 |
| Intermediate | -0.04758(+) | -0.10248(+) | -0.14486(-) | -0.19461(+) | -1671.4727 | 0.560925 |
| Product1 | -0.05547(+) | -0.09436(+) | -0.14818(-) | -0.17949(-) | -1669.1552 | 0.515745 |
| Side product | -0.04859 | -0.08056 | -0.16719 | -0.17763 | -1669.1678 | 0.514085 |
| Anthracene | -0.01046(-) | -0.05996(+) | -0.19204(-) | -0.23724(+) | -539.5305 |  |
| Tetracene | -0.01538(+) | -0.07632(+) | -0.17844(-) | -0.23242(-) | -693.1658 |  |
| Hexacene | -0.04946(+) | -0.09601(+) | -0.16209(-) | -0.20349(-) | -1000.4335 |  |
| Benzene | 0.00367(-) | 0.00360(+) | -0.24629(-) | -0.24631(+) | -232.2487 |  |
| rAziridiene | 0.06489(+) | 0.01551(+) | -0.14890(-) | -0.38050(-) | -133.8853 |  |
| Butadiene | 0.09642(-) | -0.03014(+) | -0.22734(-) | -0.31706(+) | -155.9860 |  |
| 2-butene | 0.10290(+) | 0.03683(-) | -0.23459(+) | -0.32708(-) | -157.2248 |  |

Anthracene   Tetracene 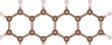  Hexacene 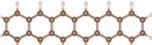

Notes: E(H$_2$) = -1.175482 Hartree
ZVE(H2) = 0.010142 Hartree

**Table S1.** Energy level and free energy of rPAMY + pentacene reaction related molecules. The unit is Hartree and + (-) represent mirror parities. The length of benzene ring makes pentacene suitable to react with rPAMY and not evaporable on Ag(100) surface.

|  | LUMO+1 | LUMO | HOMO | HOMO-1 | E(singlet) | ZVE |
| --- | --- | --- | --- | --- | --- | --- |
| rPAMY(CN) | -0.05582(-) | -0.06998(+) | -0.16054(-) | -0.23669(-) | -916.9588 | 0.268683 |
| PAMY(CN)** | -0.06725(+) | -0.10822(-) | -0.18451(-) | -0.20488(+) | -915.6108 | 0.243529 |
| PAMY(CN)-R | -0.06908 | -0.10879 | -0.15735 | -0.23283 | -916.8581 | 0.266321 |
| Product2 | -0.06365 | -0.06825 | -0.14894 | -0.20270 | -1832.7967 | 0.542854 |
| CNH | 0.02047 | 0.02047 | -0.35915 | -0.35915 | -93.4226 |  |

**Table S2.** Energy level and total energy of polyaromatic azaullazine dimer reaction related molecules. Here, the unit is Hartree and + (-) represent mirror parities.